\documentclass[11pt]{article} 
\usepackage{a4wide} 
\usepackage{latexsym}                    
\usepackage{amsfonts}                    
\usepackage{amssymb}                     
\usepackage{amsmath}                    
\usepackage{hyperref}

\def\be{\begin{equation}}
\def\ee{\end{equation}}
\def\bea{\begin{eqnarray}}
\def\eea{\end{eqnarray}}
\def\nn{\nonumber \\}

\def\vsp#1{\vspace{#1}}
\def\hsp#1{\hspace{#1}}
\def\vsp*#1{\vspace*{#1}}
\def\hsp*#1{\hspace*{#1}}

\def\part{\partial}

\def\Tr{\mbox{Tr}}
\def\Str{\mbox{Str}}
\def\STr{\mbox{STr}}

\def\mn{{\mu\nu}}

\def\makeatletter{\catcode`\@=11}
\makeatletter
\def\mathbox#1{\hbox{$\m@th#1$}}%
\def\math@ccstyles#1#2#3#4#5#6#7{{\leavevmode
      \setbox0\mathbox{#6#7}%
      \setbox2\mathbox{#4#5}%
      \dimen@ #3%
      \baselineskip\z@\lineskiplimit#1\lineskip\z@
      \vbox{\ialign{##\crcr
             \hfil \kern #2\box2 \hfil\crcr
             \noalign{\kern\dimen@}%
             \hfil\box0\hfil\crcr}}}}
\def\mathaccstyles{\math@ccstyles\maxdimen}
\def\maththroughstyles{\math@ccstyles{-\maxdimen}}
\def\unity%
 {\maththroughstyles{.45\ht0}\z@\displaystyle {\mathchar"006C}\displaystyle 1}
\begin{document}

\markboth{Ignacio Alvarez}
{Dielectric branes in non-trivial backgrounds}

\title{DIELECTRIC BRANES IN NON-TRIVIAL BACKGROUNDS}

\author{\footnotesize IGNACIO ALVAREZ ILLAN \\
Departamento de F\'{\i}sica Te\'orica y del Cosmos,\\
Universidad de Granada, 18071 Granada, Spain\\
illan@ugr.es}

\maketitle

\begin{abstract}
We present a procedure to evaluate the action for dielectric branes in non-trivial backgrounds. These backgrounds must be capable to be taken into a Kaluza-Klein form, with some non-zero wrapping factor. We derive the way this wrapping factor is gauged away. Examples of this are $AdS_5\times S^5$ and $AdS_3\times S^3\times T^4$, where we perform the construction of different stable systems, which stability relies in its dielectric character.

\end{abstract}

\section{Introduction}

Superpositions of D-branes in string theory are systems widely studied, where a field which has received extensive attention is the emergence of non-commutative geometry in this context. One phenomenon that can be seen as a  consequence of the new geometric properties of the brane superposition is the dielectric effect. The idea of this effect is that a collection of superposed neutral D$(p-2)$-branes can develop a ``dipole'' momentum under the influence of a field, expanding into a D$p$-brane. It was first explored by Emparan\cite{Emparan:1997rt} that, in the abelian theory, such an effect takes place with a collection of $N$ D$p$-branes in the presence of a R-R flux. Later, it was noted by Taylor and Van Raamsdonk\cite{Taylor:1999pr}, and Myers\cite{Myers:1999ps} that an alternative and equivalent description of this dielectric effect can be obtained in the context of the non-abelian theory, where a collection of superposed D$p$-branes in the presence of a R-R flux expand due to the fact that the transverse coordinates of the branes are matrix valued. The branes are affected by the presence of a non-commutative spacetime geometry, which produces the dipole couplings.

This non-commutative description of the dielectric effect has produced several results. Our work will follow those that have been successful in giving a microscopic description of known macroscopic systems in therms of non-commutative effects. In the context of the AdS/CFT correspondence, the non-commutative dielectric effect has been successfully employed to describe the microscopical nature of the giant gravitons\cite{Janssen:2002cf}. The giant gravitons can be constructed by a superposition of gravitational waves embedded in a non-commutative geometry, which generates an expansion of the waves into a spherical giant graviton. It has been also shown to be useful for describing black holes\cite{RodriguezGomez:2005na} and a KK monopole\cite{Janssen:2007dm}, from the same principles. Similar treatment has received the study of the baryon vertex with magnetic flux\cite{Janssen:2006sc}, where the inclusion of this flux allows to write a microscopical description of the baryon vertex from a superposition of branes.

When treating of non-commutative geometry technical difficulties arise. One of them is the ordering of the operators. Functional dependences of the operators must be ordered, but there is not a unique way to do it. One possible prescription is to take the symmetrized trace, which matches with matrix theory at first orders, as it was shown in Myer's work\cite{Myers:1999ps}. But there are other difficulties coming from this functional dependence on non-abelian operators.

In this work we present a way to deal with the problem of describing the dynamics of a superposition of branes in a non-trivial background. The non-commutative nature of the background metric is expressed through its non-abelian Taylor expansion. The key idea, developed in section 2, is to obtain systems that the background metric is reduced to a simpler one, with some components gauged away as in a gauged sigma model. In the next two sections 3 and 4, we follow with an application of the result for relevant backgrounds, like $AdS_5\times S^5$ and $AdS_3\times S^3\times T^4$, comparing the non-abelian microscopical description with the abelian macroscopic counterpart. We conclude in section 5 with a summary and perspectives of the work.

\section{The Action}

From the action proposed by Myers\cite{Myers:1999ps} for dielectric branes, we are going to develop an expression for the low-energy effective action that describes the brane dynamics, including the dielectric effect, more operative for some specific cases. We follow the Myers idea for constructing such an action, beginning with the action for superposed D9 branes, and applying T-duality to obtain the action for a collection D$p$-branes. But we focus on some specific backgrounds, where it is possible to rewrite the result of applying T-duality in a simplified way.

Reviewing the ideas that lead to Myers action, we begin with a superposition of $N$ D9-branes that completely fill the 9+1 dimensional background space. The low-energy dynamics are described by a non-abelian gauge theory with two terms: the non-abelian Born-Infeld action and a Chern-Simons term:

\[
S_{D9}=S_{BI}+S_{CS} \]
We concentrate in the Born-Infeld part:
\be
S_{BI}=-T_9\int{d^{10}\sigma \Tr \left(e^{-\phi} \sqrt{-\det{\left([g+B\right]_{ab}+\lambda F_{ab})}}\right)}
	\label{bi}
\ee
In this action we will take static gauge, forcing the ten worldvolume coordinates $\sigma^a$ to coincide with with the ten target space coordinates $x^\mu$. Also the Born-Infeld form will be non-abelian:
\[
F_{ab}=2\partial_{[a}A_{b]}+i[A_a,A_b]	\ \ \ \ \ A_a=A_a^I T_I
\]
Where $T_I$ are the generators of the $U(N)$ gauge symmetry that is contained in the worldvolume of the $N$ D9 branes, 

The background metric $g$ together with the Kalb-Ramond field $B$ constitute a unique background field $E$, under the redefinition $E\equiv g+B$, which is useful for further calculations. 

Let's apply T-duality to the action (\ref{bi}). The T-duality rules are\cite{Shapere:1988zv,Giveon:1994fu}:
\be
\begin{split}
\tilde{E}_{ab}\rightarrow &\ E_{ab}-E_{ai}E^{ij}E_{jb}\\
\tilde{E}_{ai}\rightarrow &\ E_{aj}E^{jk}	\\
\tilde{E}_{ij}\rightarrow &\ E^{ij}\\	
e^{-\phi}     \rightarrow &\ e^{-\phi}/\sqrt{\det{E^{ij}}}	\label{tduales}
\end{split}\ee
where $E^{ik}E_{kj}=E_{jk}E^{ki}=\delta^i_j$. Implicitly, the previous equations summarize the application of T-duality not only through one isometric direction, but iterated $9-p$ times through the $i,j=p+1,...,9$ directions. The remaining coordinates, under which we do not apply T-duality, are labeled with indices from the beginning of the alphabet $a,b$. 

For the BI field $F$, it is also possible to summarize the result of the same application of T-duality\cite{Myers:1999ps}:

\begin{equation}\begin{split}
\tilde{F}_{ab}\rightarrow &F_{ab} 	\\ \tilde{F}_{ai}\rightarrow &\ D_a \Phi^i \\ \tilde{F}_{ij}\rightarrow &\ i[\Phi^i,\Phi^j]
	\label{tdualf}
\end{split}\end{equation}
where we have used the covariant derivative:
\be\label{cd}
D_a \Phi^i=\partial_a \Phi^i+i[A_a,\Phi^i]
\ee

The component of the potential vector $A_p$ wrapped in the direction $p$ transform under T-duality to the non-abelian scalar $\Phi^p$. The brane dynamics are described by these $9-p$ scalars $\Phi^i$ that represent the transverse displacements of the branes. It is possible to identify this scalars with a non-commutative 'coordinate' system, under $X^i=2\pi\alpha^\prime\Phi^i$. Under such identification, the transverse space of the brane will have a non-commutative geometric nature. 

Putting together this T-duality transformations in the action (\ref{bi}), and taking into account that $\lambda=2\pi\alpha^\prime$, the magnitude under the square root $D=-\det{(\left[g+B\right]_{ab}+\lambda F_{ab})}$ transforms as:

\begin{equation}\label{tildeddd}
D=-\det
\left( \begin{array}{cc}
E_{ab} - E_{ai} E^{ij} E_{jb} + \lambda F_{ab} & E_{ak}E^{kj} + \lambda D_a \Phi^j  \\
-E^{ik}E_{kb} - \lambda D_b \Phi^i  & E^{ij} + i\lambda [\Phi^i, \Phi^j]   
\end{array} \right)
\end{equation}

Here we stop the review. One can evaluate the determinant in the above expression\footnote{If doing so, the result obtained is the action proposed by Myers:
\[
S_{BI} = -T_p\int d^{p+1}\sigma \STr\left(e^{-\phi} \sqrt{-\det (P [E_{ab} + E_{ai}(Q^{-1}-\delta )^{ij}E_{jb}] +\lambda F_{ab})\det(Q^{i}_{j})}\right)
\]
where the pullback is computed with covariant derivatives defined in (\ref{cd}), the transformation of the dilaton has been taken into account, the $x^i$ coordinates have been integrated to give the prefactor $T_p$, and:

\[
Q^i_j\equiv \delta ^i_j+ i\lambda[\Phi^i,\Phi^k]E_{kj}	
\]}, but first we want to consider only some specific backgrounds. We focus our interest in configurations where the background metric can be put into a Kaluza-Klein form with some wrapping factor\cite{Duff:1986hr}
, written in the following way:

\be\label{fibration}
\begin{split} ds^{2}=&g_{ab} dy^a dy^b+\tilde{g}_{ij} dx^{i}dx^{j}+k(dy^p+A_idx^i)^{2} =\\ 
=&g_{ab} dy^{a} dy^{b}+{g}_{ij} dx^{i}dx^{j}+k(dx^p)^2+2kA_idx^idx^p
\end{split}
\ee where
\be\label{tildg}\begin{split}
\tilde{g}_{ij} = &{g}_{ij} -kA_iA_j \\
i,j=&p+1,\ldots,9 \\ a,b=&(0,1,\ldots,p-1)
\end{split}\ee

We make a distinction between the $y$'s, with subindices from the alphabet beginning, and the $x$'s coordinates, with subindices $i,j,...$, in such a way that there is a correspondence between the notation of the T-duality transformations (\ref{tduales}) and (\ref{tdualf}), and the metric.

Once the background metric we are interested in is determined, we will make some extra simplifying assumptions to evaluate the determinant in (\ref{tildeddd}). We will take $F_{ab}=0$, and $B_{\mu\nu}=0$, leaving $g_\mn$ as the only non-zero massless background field. The background metric in presence of the brane pile system will be a functional of the non-abelian transverse scalars, given by the non-abelian Taylor expansion\cite{Hull:1997jc,Garousi:1998fg}:
\[
g_\mn=\exp{[\lambda \Phi^i\partial_i]}g_\mn^0(\sigma^a,x^i)|_{x^i=0}
\]

The determinant (\ref{tildeddd}) we are interested in, will only depend on the metric for this configuration: 

\begin{displaymath}
D=-\det
\left( \begin{array}{ccc}
g_{ab} & 0 & \lambda \partial_b\Phi^j \\
0 & k - k^{2}A_{i} g^{ij} A_{j}  & kA_{k} g^{kj}+ \lambda \partial_p \Phi^j\\
-\lambda \partial_a\Phi^i & -kg^{ik} A_{k} -\lambda \partial_p \Phi^i  & g^{ij} + i \lambda [\Phi^i, \Phi^j]   
\end{array} \right)
\end{displaymath}

Making use of the determinant properties, it is straightforward to calculate this determinant, to obtain:

\[
D=-\det (g_{ab})\det(g^{il})\det(\tilde{Q}_{l}^{j}+\lambda^2\tilde{g}_{kl}g^{ab}\partial_a\Phi^l\partial_b\Phi^j)) 
\]
where $\tilde{Q}$ is:
\[
\tilde{Q}_l^j=\delta_l^j + i\lambda \tilde{g}_{lk} [\Phi^k, \Phi^j]
\]
Recall that in the definition of $\tilde{Q}$ appears $\tilde{g}$ instead of $g$, which are related by (\ref{tildg}). It is interesting to note that $\tilde{g}$ resembles the metric of a \textsl{gauge-sigma model}, where some metric components are gauged away. 

Making use of these results, the transformed BI part of the D9-action under T-duality, considering also the dilaton transformation, yields to:

\begin{equation}\label{ao} 
S_{BIN} = -T_p\int d^{p+1}\sigma \STr\left(e^{-\phi}\sqrt{g} \sqrt{\det {(\tilde{Q}_k^j+\lambda^2\tilde{g}_{kl}g^{ab}\partial_a\Phi^l\partial_b\Phi^j)} }\right) 
\end{equation}
where 
\begin{equation}\label{sqrtg}
\sqrt{g}=(-k\det{g}_{ab})^\frac{1}{2}
\end{equation}
This part of the action describes the low-energy dynamics of $N$ superposed D$p$ branes in a non-trivial background $g_\mn$ that fulfill (\ref{fibration}). With static gauge choice, worldvolume coordinates $\sigma^a$ span the worldvolume with metric $g_{ab}$. In the other hand, the transverse space with metric $g_{ij}$ will turn a non-commutative space, described by the adjoint scalars $\Phi^i$. They can be interpreted as non-commutative coordinates, under the identification $X^i=2\pi\alpha^\prime\Phi^i$, that span the non-commutative subspace transverse to the branes. 

In the resulting action, the forms appearing on it are no longer pulled back as in Myers action. Instead, there is a sum of two terms. Focusing in the second term $\lambda^2\tilde{g}_{kl}g^{ab}\partial_a\Phi^l\partial_b\Phi^j$, it can be seen as a gauged sigma model pull-back to the metric, as it has been used to describe KK monopoles\cite{Bergshoeff:1997gy} and others\cite{Bergshoeff:1997ak}.

The interesting thing about this result is that the action is independent of the wrapping factor $A_i$ of the KK metric \ref{fibration}, in such a way that the effective transverse space is described by $\tilde{g}_{ij}$ and not by $g_{ij}$. 
There are known examples in which this simplification takes place\cite{Janssen:2006sc} and allows to write down a non-commutative ansatz for the transverse space.
We follow with an application of this result.

\section{The $AdS_5\times S^5$ case}

The background metric in $AdS_5\times S^5$ satisfies the condition (\ref{fibration}) by writing the $S^3$ part in $AdS_5$ as a Hopf fibration:

\begin{equation}
	ds^2=-\left(1+\frac{r^2}{L^2}\right)dt^2+\left(1+\frac{r^2}{L^2}\right)^{-1}dr^2+\frac{r^2}{4}(d\theta^2+\sin\theta^2d\psi^2+(d\chi+\cos\theta d\psi)^2)+L^2d\Omega_5 ^2
	\label{background1}
\end{equation}
where the base $S^2$ of the Hopf fibering is described by the angles $\theta$ and $\psi$. It will be useful for later calculations the use of Cartesian coordinates:
\be\begin{split}\label{coor}
X^1=&R\sin\theta\cos\psi \\
X^2=&R\sin\theta\sin\psi \\
X^3=&R\cos\theta
\end{split}\ee

This background has a self dual five form $F^{(5)}=\ast F^{(5)}$, determined by:

\[
F^{(5)}=\sqrt{g_{S^5}}
\]
Now one can introduce in this background a superposition of $N$ D1-branes. The brane pile will wrap the $\chi$ direction (the $S^1$ fiber of the $S^3)$ and will have $\tau=t$, leaving as transverse space to them the direct product of the $S^5$ part, times the subspace spanned by $r$, and the $X^i$ with $i=1,2,3$. We focus on configurations in which $r$ is constant, restricting us to brane piles that conform spherical states in $AdS_5$. It is appropriate to describe this configuration by the action obtained in (\ref{ao}), being necessary to include also Chern-Simons couplings. The full action will take the form:
\[
S=S_{BIN}+S_{CS}
\]
For this system, the relevant Chern-Simons coupling is\cite{Myers:1999ps}:
\[
S_{CS}=i\lambda T_1\int d\tau d\chi\Str( P[(i_{\Phi}i_{\Phi})C^{(4)}])
\]
that generates a 'dipole' coupling and induce the dielectric effect in our pile of branes, where we follow the notation:
\[
i_{\Phi}i_{\Phi}C^{(n)}=\frac{1}{2}[\Phi^i,\Phi^j]C^{(n)}_{ij\ a_1...a_{n-2}}
\]

The self-dual five form $F^{(5)}$ is the field-strength of the 4-form potential $F^{(5)}=dC^{(4)}$. The only $C^{(4)}$ components which are relevant in this configuration, using the above coordinates, can be expressed as:
\[
C^{(4)}_{ij\chi t}=-\frac{r^4}{8R^3L}\epsilon_{ijk} X^k
\]
In order to describe the non-commutative nature of the transverse space we identify $X^i=\lambda\Phi^i$. It is implicit that we consider abelian all the other transverse coordinates but $X^i$ ($i=1,2,3$). This consideration is supported by the fact that commuting matrices $\Phi^i$ is a possible configuration together with the fact that the transverse space can be splited into direct product parts.  

The identification made, allows us to give a complete description of the system, putting together the non-commutative BI part and the CS part. The full action is:
\[
S=S_{BIN}+S_{CS}=
-T_1\int d\tau d\chi \STr\left(\sqrt{g} \sqrt{\det {(\tilde{Q}_k^j+\lambda^2\tilde{g}_{kl}g^{ab}\partial_a\Phi^l\partial_b\Phi^j)} }-i\lambda  P[(i_{\Phi}i_{\Phi})C^{(4)}]\right)
\]
where we have made use of the fact that there is no dilaton in this background. Considering the low-energy regime, we expand the square root of the previous expression in terms of $\lambda$, obtaining a kinetic term proportional to $(\partial \Phi)^2$, plus a potential:

\[
V(\Phi)=\STr\left(-4\pi T_1\sqrt{g}\left(\unity-\frac{\lambda^{2}}{4}\tilde{g}_{ij}\tilde{g}_{kl}[\Phi^j,\Phi^k][\Phi^i,\Phi^l]\right)+\frac{i\lambda^2 T_1r^4}{4R^3L}\epsilon_{ijk} [\Phi^i,\Phi^j]\Phi^k\right)+{\cal O}(\lambda ^3)
\]
Taking advantage of the use of (\ref{ao}) to describe this system, a simplification occurs in the description of the transverse metric in the coordinates (\ref{coor}). The non-abelian Taylor expansion of the metric $\tilde{g}$ functional of $\Phi$, only has a constant term:
\[
\tilde{g}_{ij}=\frac{r^2}{4R^2}\delta_{ij}
\]
Finally, the branes will be under the influence of the potential:
\[
V(\Phi)=\STr\left(-4 \pi T_1\sqrt{g}\left(\unity-\lambda^{2}\frac{r^4}{64R^4}([\Phi^i,\Phi^j])^2\right)+\frac{i\lambda^2 T_1r^4}{4R^3L}\epsilon_{ijk} [\Phi^i,\Phi^j]\Phi^k\right)+{\cal O}(\lambda ^3)
\]

Looking for a solution to $\delta V/\delta \Phi^i=0$, is the same as looking for a stable configuration of the system. For the system to have a stable configuration, the equation that $\Phi^i$ must satisfy is:
\be\label{min}
\frac{\sqrt{g}}{4}[[\Phi^i,\Phi^j],\Phi^j]+i\frac{3R}{L} \epsilon_{ijk}[\Phi^j,\Phi^k]=0
\ee
The particular solution that is consistent with (\ref{coor}) is the fuzzy sphere\footnote{$r=0$ and the commuting coordinates $[\Phi^i,\Phi^j]=0$ are also solutions to $\delta V/\delta \Phi^i=0$. }\cite{Madore:1991bw}. Its description, through $N\times N$ matrices, tends to the 2-sphere in the large $N$ limit. The construction comes from the identification of the non-commutative coordinates $X^i$ with certain $N\times N$ matrices  $J^i$, which conform a irreducible representation of the $SU(2)$ algebra. We will describe it taking the ansatz:
\be\label{casimir}\begin{split}
X^i=\lambda\Phi^i=&\frac{ R\:J^i}{\sqrt{N^2-1}} \\
(X^1)^2+(X^2)^2+(X^3)^2=&R^2\unity
\end{split}\ee
where $J^i,\ i=1,2,3$  matrices are the generators of irreducible representations of $SU(2)$, $[J^i,J^j]=2i\epsilon^{ijk}J^k$, and the non-commutative coordinates satisfy the Casimir relation necessary to reproduce the abelian relation. In the large number of branes limit, the commutator of the non-commutative coordinates vanishes, reproducing the abelian behavior. Substituting this ansatz into (\ref{min}), and taking into account that from (\ref{sqrtg}) and (\ref{background1}) $\sqrt{g}=\frac{r^2}{4}\sqrt{1+\frac{r^2}{L^2}}$, we obtain that there exist a stable configuration for the system if:

\be\label{rad}
r=\frac{L}{\sqrt{2}} \sqrt{\sqrt{1+\frac{2(12\lambda)^2(N^2-1)}{L^4}}-1}
\ee

For this value of $r$, the potential reaches its minimum, which value is less than for commuting matrices: the configuration is stable. This represents an example of how the dielectric effect occurs in a non-trivial background. The D1-branes expand into a non-commutative D3-brane and configure an stable system, which stability is not dynamic but relies in its dielectric character.

Once we know how a solution must be, we can insert our ansatz (\ref{casimir}) for the non-commutative coordinates into the general action. After substituting and taking the symetrized trace, we get the Taylor expansion first terms of the potential:

\be\label{prev}
V=-4 \pi T_1N\left(\sqrt{\frac{r^2}{4}\left(1+\frac{r^2}{L^2}\right)\left(1+\frac{r^4}{4\lambda^{2}(N^2-1)}\right)}- \frac{r^4}{\lambda L(N^2-1)^{1/2}}\right)
\ee
where the square root is to be expanded in powers of $\frac{r^4}{4\lambda^{2}(N^2-1)}$, which must remain small in order to give a consistent result.
This potential has a minimum which is reached for a value of the radius that can be approximated by the previous one (\ref{rad}). 

\subsection*{Macroscopic description}

We can give a macroscopic description of the non-commutative superposition of branes system in terms of an static abelian D3-brane wrapping the $S^3\subset S^5$ part of the $AdS_5\times S^5$ background. Consider a setting conformed by a D3-brane with worldvolume coordinates $\tau, \theta,\psi,\chi$ from (\ref{background1}), subject to the influence of a electromagnetic flux. The full low-energy dynamics will be described by a Born-Infeld action term plus a Chern-Simons of the form:

\begin{equation}
S=S_{BI}+S_{CS}=-T_3\int_{\mathbb{R}\times S^3}{e^{-\phi}\sqrt{-\det\left(P[g]+\lambda F\right)}}-\;T_3\int_{\mathbb{R}\times S^3}{P[C^{(4)}]}
	\label{eqcs}
\end{equation}

The background has a non-zero RR 4-form inducing a Chern-Simons coupling. The absence of the square root $\hat{\cal A}$-genus terms of higher dimension as the first constant term is justified by the fact that $C^{(0)}$ is zero in this background, not allowing to form the couplings for the four dimensional worldvolume. One should also expect a term of the form $\int_{\mathbb{R}\times S^3} P[C^{(2)}]\wedge F$, that would dissolve $n$ units of D1-brane charge in the worldvolume of the D3 brane system, by introducing a field strength such $\int_{S^2}F=2\pi n$ combined with the electric components of $C^{(2)}$. But, because $C^{(2)}=0$ in this background, this term would be also absent. Nevertheless we will include a field that fulfill this condition, i.e., one such:

\begin{equation}
F_{\theta \psi}=-\frac{n}{2} \sin\theta 
	\label{fluxx}
\end{equation} 

Introducing such a field, in spite of the desired Chern-Simons coupling term absence, will allow us to relate this macroscopic description with the previous. The D3-brane, which wraps the $S^3$ part of $S^5$, is not topological stable. But due to the flux presence, there will be an equilibrium that will generate the macroscopic description of the previous stable state. This equilibrium has not a dynamic character, but is due to $F$ field action into the static  charged D3-brane.

The action (\ref{eqcs}) in the background (\ref{background1}) fixes the Lagrangian density ${\cal L}$ for a static trial D3-brane, with worldvolume coordinates $\tau, \theta,\psi,\chi$, and subject to the influence of a RR flux (\ref{fluxx}). For this static solution, the potential energy is given by:
\[
V=-T_3\int d\theta d\psi d\chi {\cal L}=-T_3\int d\theta d\psi d\chi\left(\sqrt{\frac{r^2}{4}\left(1 + \frac{r^2}{L^2}\right)\left(\frac{\lambda^2n^2}{4} + \frac{r^4}{4^2}\right)\sin^2\theta} - \frac{r^4}{4L}\sin\theta\right)
\] 
After integration, and considering that $T_p=2\pi\lambda T_{p+2}$, the outcome is:

\[
V=-4\pi T_1n \left(\sqrt{\frac{r^2}{4}\left(1 + \frac{r^2}{L^2}\right)\left(1 + \frac{r^4}{4\lambda^2n^2}\right)} - \frac{r^4}{2n\lambda L}\right)
\] 

In this way, we recover the action that we obtained previously for the microscopic description (\ref{prev}), provided that we take the large number of branes $N$ limit. In the microscopic case we made some approximations, in such a way that the potential (\ref{prev}) was only valid for $r^4<<\lambda^2(N^2-1)$. Since the measure of an area element of the fuzzy sphere is $4\pi r^2/N$ \cite{Kabat:1997im,Rey:1997iq}, our previous results are only valid for small fuzzy-sphere areas, in terms of the string-scale, which is consistent with the low-energy approximation.

\section{The $AdS_3\times S^3 \times T^4$ case}

The $AdS_3\times S^3 \times T^4$ background appears as the near horizon limit of the D1-D5 brane intersection\cite{Maldacena:1997re,Maldacena:1998bw}. The $AdS_3\times S^3 \times T^4$ background metric satisfies the condition (\ref{fibration}), by writing the $S^3$ part as a Hopf fibration:

\begin{equation}\label{background}	\begin{split}ds^2=&-\left(1+\frac{r^2}{L^2}\right)dt^2+\left(1+\frac{r^2}{L^2}\right)^{-1}dr^2+r^2d\varphi^2+\\
+&\frac{L^2}{4}(d\theta^2+\sin\theta^2d\psi^2+(d\chi+\cos\theta d\psi)^2)+\mathcal{R}^2dy_m^2
\end{split}
\end{equation}

The factors and the remaining background fields (the dilaton $\phi$ and the two-form R-R potential $C^{(2)}$) are determined by: 
\begin{equation}
e^\phi=\mathcal{R}^2=\sqrt{Q_1/Q_5}\ \ \ \  L^2=\sqrt{Q_1Q_5}\ \ \ C_{t \varphi}^{(2)}=-\frac{Q_5}{L^3}r^2\ \ \ C_{ \psi \chi}^{(2)}=Q_5 \cos{\theta}
\end{equation}
where $Q_1$ and $Q_5$ are the charges of the D1 and D5 branes that conform the D1-D5 intersection system.

We will introduce a pile of probe branes in this background, in order to apply the previous results. Our system will be formed by a superposition of $N$ D1-branes, in spite of the fact that we could have chosen other dimensionalities: transforming under T-duality our D1-branes through the directions of the $T^4$, it is possible to obtain equivalent systems of D2, D3, D4, or D5 branes.

The D1-branes of the system will be wrapped in the $S^1$ part of the $S^3$ background taking also static gauge, that is, the worldvolume coordinates will be $(t,\chi)$. We will consider only static configurations in which $r$ and $\varphi$ are constant. The subspace transverse to the branes will be conformed by the direct product of the $S^2$ part of the Hopf fibration and the remaining $AdS_3$ part, times the $T^4$ part. 

In the presence of the superposed branes, this transverse space will turn into a non-commutative space. For later calculations, it will be useful to describe the $S^2$ part of the Hopf fibration in the background metric in Cartesian coordinates $X^i$ ($i=1,2,3$), in such a way that the Cartesian coordinates satisfy $X_iX^i=R^2$, in the same fashion as (\ref{coor}).

Once the configuration of the branes is determined, we focus our interest on the action that describes the low-energy sector of this system. It will only have a non-commutative Born-Infeld term, because the Chern-Simons couplings will be zero in this configuration. Therefore, the system can be described by (\ref{ao}):
\[
S_{BIN} = -T_1\int dt d\chi \STr\left(e^{-\phi}\sqrt{g} \sqrt{\det {(\tilde{Q}_k^j+\lambda^2\tilde{g}_{kl}g^{ab}\partial_a\Phi^l\partial_b\Phi^j)} }\right) 
\]

The transverse space to the brane pile has a non-commutative nature, which we take into account under the identification\footnote{we are assuming that any other coordinate but $\Phi^i$ behaves as an abelian one} $X^i=\lambda \Phi^i$.  Now the relevant transverse metric in cartesian coordinates is:
\[
\tilde{g}_{ij}=\frac{L^2}{4R^2}\delta_{ij}
\]
The functional dependence of the transverse metric on $\Phi^i$ is simply given by the first constant term of the non-abelian Taylor expansion, for this particular configuration. Substituting in the action:
\[
S_{BIN} = -T_1\sqrt{\frac{Q_5}{Q_1}}\int dt d\chi \STr\left(\sqrt{g} \sqrt{\det {(\delta_i^j+\frac{i\lambda L^2}{4R^2}[\Phi_i,\Phi^j]+\lambda^2\frac{L^2}{4R^2}g^{ab}\partial_a\Phi_i\partial_b\Phi^j)} }\right) 
\]

The expansion in $\lambda$ of the square root generates two terms, a kinetic term proportional to $(\partial \Phi)^2$ plus a potential:

\be\label{pot}
V(\Phi)=4 \pi T_1\sqrt{\frac{Q_5}{Q_1}}\STr\left(\sqrt{g}\left(\unity-\lambda^{2}\frac{L^4}{4^3R^4}([\Phi^i,\Phi^j])^2]\right)\right)+{\cal O}(\lambda ^3)
\ee

In order to recover the relation $X_iX^i=R^2$ a further ansatz in the $X^i=\lambda\Phi^i$ must be taken. 
One known form of describing the 2-sphere with non-commutative coordinates is its counterpart fuzzy 2-sphere\cite{Madore:1991bw}:
 
\be\label{casmir}\begin{split}
X^i=&\frac{R\:J^i}{\sqrt{N^2-1}} \\ \nn
\left[J^i,J^j\right]=&2i\epsilon^{ijk}J^k
\end{split}\ee
and with casimir
\[
(X^1)^2+(X^2)^2+(X^3)^2=R^2\unity
\]
With this ansatz, it is straightforward to evaluate the potential (\ref{pot}), taking under consideration that:
\begin{equation}\label{detq}
\det \tilde{Q}=\unity+\frac{L^4}{4\lambda^2(N^2-1)R^2}X^2
\end{equation}
so, after taking the symmetrized trace:
\begin{equation}\label{energy1}
V=-\frac{4 \pi T_1Q_5N}{L^2}\sqrt{\frac{L^2}{4}\left(1+\frac{r^2}{L^2} \right)\left(1+\frac{L^4}{4\lambda^2(N^2-1)}\right)}
\end{equation}
where the previous square root function is an approximation, only valid for $L^4<<4\lambda^2(N^2-1)$. In this case, it is possible to expand the square root, keeping only the first Taylor expansion terms, which recover the direct substitution of (\ref{detq}) into the potential after taking the symmetrized trace. 

The construction we have presented is an stable one, where the non-abelian superposition of D1-branes expand in the non-commutative geometry of the transverse space, expanding to the $S^3$. This guarantees the topological stability of this static system, which energy can be obtained directly as minus the potential.


\subsection*{Macroscopic description}

We can give a macroscopic description of the system in terms of a static D3-brane wrapped in the background $AdS_3\times S^3\times T^4$, in the presence of a Born-Infeld vector flux. 

The background has a non-zero RR 2-form inducing a coupling in the Chern-Simons part of the action: 
\[
S=S_{BI}+S_{CS}=-T_3\int_{\mathbb{R}\times S^3}{e^{-\phi}\sqrt{-\det\left(P[g]+\lambda F\right)}}-\;T_3\int_{\mathbb{R}\times S^3}{P[C^{(2)}]\wedge F}
\] 
We will explore which choice of the BI vector reproduces our non-abelian result. Firstly we will choose $F$ in such a way that it would dissolve $n$ units of magnetic D1 charge in the D3-brane. To be so, it must be of the form: $F=dA$ with $A=n\cos\theta d\psi$. This gauge choice of the BI vector allows to rewrite the Chern-Simons coupling in the following way:

\[
S_{CS}=T_3\lambda\int_{\mathbb{R}\times S^3}{P[C^{(2)}]\wedge F}=2\pi n T_3 \lambda\int_{\mathbb{R}\times S^1}{P[C^{(2)}]}=n T_1\int_{\mathbb{R}\times S^1}{P[C^{(2)}]}
\]
where we have made use of $T_p = 2\pi\lambda T_{p+2}$ and we have taken into account that $\int_{S^2} F = 2\pi n$, leading to the same CS coupling as the one for $n$ charged D1-branes. 

We will use this choice of $F$, but for this concrete configuration for the D3-brane, there will not be any CS coupling. For reproducing the previous results, the probe D3-brane must be wrapped in the $S^3$ part of the background, considering also static gauge. This configuration is topological stable, due to the $S^3$ topology. This equilibrium will not be broken by our choice of the BI vector, because the electric components of $C^{(2)}$, necessaries to couple to $F$ in a CS coupling term form, are zero in the $S^3$ part of the background, where our probe brane is wrapped. The same occurs for $C^{(0)}$, not being possible to form gravitational couplings depending on the expansion of the square root of the $\hat{\cal A}$-genus of the manifold. 

There will be only a BI term with non-zero $F$, which fixes the Lagrange density ${\cal L}$. Consequently, the static brane is under the influence of the potential:
\[
V=\int  d\theta d\psi d\chi {\cal L}=-\frac{T_3Q_5\lambda n}{L^2}\int  d\theta d\psi d\chi   \sqrt{\frac{L^2}{4}\left(1+\frac{r^2}{L^2} \right)\sin^2{\theta}\left(1+\frac{L^4}{4n^2\lambda^2}\right)}
\]
which, after integration, leads to:
\[
V=-\frac{4 \pi T_1Q_5n}{L^2}  \sqrt{\frac{L^2}{4}\left(1+\frac{r^2}{L^2}\right)\left(1+\frac{L^4}{4n^2\lambda^2}\right)}
\] 
Comparing this result with (\ref{energy1}) we find a exact agreement. We see that these configurations with magnetic flux have a dual description in terms of expanding microscopical non-abelian branes. Both match precisely in the limit $L^4<<4\lambda^2(N^2-1)$, as stated below (\ref{energy1}).

However another consistent choice for $F$ is possible. If we take it to be a electric flux and not a magnetic one, the construction we arrive bears resemblance with the baryon vertex\cite{Witten:1998xy}, but in a different background. An electric BI field would induce $N$ units of electric charge on the brane:

\[
S_{CS}=T_3\lambda\int_{\mathbb{R}\times S^3}{P[C^{(2)}]\wedge F}= T_3 \lambda\int_{\mathbb{R}\times S^3}{G^{(3)}}A_t=N T_1\int_{\mathbb{R}}A_t
\]

Provided that an electric choice for the BI vector would be $A=\sin\theta dt$. In this case, the CS coupling will be non-zero, giving charge to the spherical brane. But due to the fact that the D3-brane wraps a closed manifold, it can not have any global charge, which tell us that we must include something more in our action. What we must include is the action of $N$ open strings coming from the AdS part, and ending on the D3-brane, producing in its worldvolume the necessary opposite charge. This would be represented by a Nambu-Goto term with a boundary term, in such a way that the complete action is:

\[
S=S_{BI}+S_{CS}+S_{NG}+S_{\partial M}=S_{BI}+N T_1\int_{\mathbb{R}}A_t+S_{NG}-N T_1\int_{\mathbb{R}}A_t
\]

The energy of this system consist now in two parts. The energy corresponding to the D3 branes with the previous gauge choice for $A_t$ would also match the microscopic one of the previous chapter, which states a manifest parallelism with the work\cite{Janssen:2006sc}. It would be interesting to explore the connections with the CFT counterpart of this system, beyond the scope of this work.
\section{Conclusions}

We have presented an alternative and simplified way to deal with systems of superposed branes. We have restricted ourselves to systems in which the background metric can be brought into a KK form. To this kind of backgrounds belong the fibrations ones, such as $S^3$ seen as a Hopf fibration. We have also presented relevant examples of backgrounds that have $S^3$ as a subspace, where our results do apply. Making use of configurations where the branes are wrapped in the fibre direction $S^1$, the transverse space to the branes is reduced to a non-commutative $S^2$. Roughly speaking, the relevant subspace $S^3$ turns effectively into a $S^1\times S^2_{fuzzy}$. Technical difficulties are overcome by the use of (\ref{ao}) instead of Myers action for dielectric branes, and two stable solutions are constructed with a perfect agreement found between our description and the macroscopic one, provided that a low-energy approximation is made, together with a small fuzzy-sphere area approximation is taken. We have left opened the possibility to explore the CFT duals of these two solutions in the $AdS_5\times S^5$ and $AdS_3\times S^3\times T^4$ backgrounds, where we expect a connection with the stringly exclusion principle for the first one and some kind of baryon vertex counterpart for the second one.
\section*{Acknowledgments}
We wish to thank Rafael Hernandez for the encouragement, useful help and discussions.
The work of I.A.~is partly supported by the MICINN under the PETRI DENCLASES (PET2006-0253), TEC2008-02113, NAPOLEON (TEC2007-68030-C02-01) and HD2008-0029 projects and the Consejería de Innovaci\'{o}n, Ciencia y Empresa (Junta de Andalucía, Spain) under the Excellence Project TIC-02566.. 



\end{document}